 \documentclass[final,5p,times,twocolumn]{elsarticle}

\usepackage{amssymb}
\usepackage{lipsum}
\usepackage{amsmath}

\usepackage{xcolor}

\journal{Annals of Physics}

\begin{document}

\begin{frontmatter}



\title{Are Padé approximants suitable for modelling shape functions of traversable wormholes?}


\author[first]{Jonathan Alves Rebouças}
\affiliation[first]{organization={State University of the Ceará (UECE), Federal Institute of Education, Science and Technology of the Ceará (IFCE)},
            addressline={}, 
            city={Iguatu},
            postcode={}, 
            state={Ceará},
            country={Brazil}}
\author[second]{Celio Rodrigues Muniz}
\affiliation[second]{organization={State University of the Ceará (UECE)},
            addressline={}, 
            city={Iguatu},
            postcode={}, 
            state={Ceará},
            country={Brazil}}

\begin{abstract}
This study investigates the applicability of Padé approximants in constructing suitable shape functions for traversable wormholes, emphasizing their ability to satisfy essential geometric constraints. By analyzing low-order Padé approximants, we demonstrate their effectiveness in transforming inadequate shape functions into physically consistent candidates, while inherently fulfilling critical criteria such as asymptotic flatness, flare-out conditions, and throat regularity. Specific parameter restrictions are established to ensure compliance with these constraints; for instance, low-order rational approximations help to avoid artificial singularities and maintain asymptotic behavior when derivative conditions at the throat are controlled. In contrast, high-order Padé approximants introduce challenges, including spurious poles within the physical domain, which disrupt geometric requirements. Our findings highlight that low-order Padé approximants provide a robust framework for simplifying complex shape functions into analytically tractable forms, balancing mathematical flexibility with physical feasibility. This work underscores their potential as a systematic tool in traversable wormhole modeling, while cautioning against unphysical artifacts in higher-order approximations.
\end{abstract}




\end{frontmatter}




\section{Introduction}

The investigation of wormhole geometries has evolved from a predominantly theoretical curiosity into a significant discipline within gravitational physics \cite{morris1988wormholes,saleem2023traversable}. Initial analyses regarded wormholes as specific solutions to the Einstein field equations \cite{flamm1916comments,einstein1935particle}; however, contemporary research has considerably broadened their scope \cite{guerrero2022light,olmo2015geodesic,dai2019observing}. High-performance numerical relativity has enabled rigorous simulations of wormhole spacetimes, clarifying their stability under both linear and non-linear perturbations, interactions with scalar and electromagnetic fields, and the dynamical evolution of their throats \cite{jafferis2022traversable}. Concurrently, observational strategies aimed at identifying characteristic signatures—such as anomalous gravitational lensing patterns, distinctive echoes in gravitational-wave signals, or subtle deviations in electromagnetic spectra—have opened new avenues for empirical investigation \cite{jusufi2018gravitational,dai2019observing,de2020general,simonetti2021sensitive}. Furthermore, wormhole models serve as critical testbeds for probing fundamental issues related to causality, topology changes, and horizon thermodynamics \cite{rani2022cosmographic}. By providing theoretical platforms for extreme gravitational regimes, they contribute to ongoing efforts in quantum gravity, enhance our understanding of black hole interior structures, and clarify the relationship between spacetime topology and matter-field configurations \cite{gonzalez2010graphene}. Collectively, these developments illustrate that wormholes have transformed into indispensable constructs in modern gravitational research, necessitating increasingly sophisticated analyses of their intrinsic properties and potential observables.

In addition to these achievements, modified theories of gravity have generated new perspectives on wormhole configurations , frequently reclassifying them as non-standard solutions that transcend the constraints of classical general relativity \cite{nojiri2007econf}. Driven by unresolved phenomena such as cosmic inflation, late-time acceleration, and the dark-energy enigma, researchers have formulated extensions to the Einstein–Hilbert action \cite{starobinsky1980new,copeland2006dynamics,capozziello2011extended}. These extensions include nonlinear functions of the Ricci scalar ($f(R)$ gravity), explicit couplings between curvature invariants and the stress–energy tensor ($f(R,T)$ models), torsion-based frameworks ($f(T)$ gravity), and non-metricity formulations ($f(Q)$ gravity), among others \cite{brans1961mach, bertolami2007extra,houndjo2012reconstruction}. Within these generalized contexts, the modified field equations enable a broader range of geometries, with wormholes often sustained by effective stress–energy contributions arising from curvature corrections or nonminimal couplings \cite{naz2023evolving,nazavari2023wormhole,solanki2023wormhole,tayde2023wormhole}. In such models, requirements that traditionally demanded exotic matter—specifically, violations of the null energy condition or negative energy densities—can be reinterpreted as emerging from geometric modifications rather than from unphysical matter sources \cite{saleem2023wormhole}. This reconceptualization broadens the theoretical plausibility of traversable wormholes (TWH) and enhances our understanding of how gravitation might naturally accommodate nontrivial topologies \cite{azizi2013wormhole}. Comparative studies across diverse gravitational frameworks provide critical benchmarks for identifying theories that inherently support wormhole solutions and for assessing their consistency with established physical principles \cite{naz2023evolving}.

A fundamental element in any traversable wormhole construction is the shape function, which delineates the radial metric profile of the throat and ensures compliance with essential geometric criteria. A suitably defined shape function guarantees the absence of event horizons or curvature singularities at the throat, satisfies the flare-out condition to maintain throat openness, and moderates tidal accelerations experienced by traversing observers \cite{morris1988wormholes,flamm1916comments,visser1995lorentzian}. Historically, many investigations have selected analytic shape functions based on heuristic convenience or to simplify the resolution of the modified field equations \cite{mishra2020traversable,ahmed2022existence,naseer2023constructing}. These choices often ensured boundary and regularity conditions by construction. While such approaches have produced instructive examples, they raise questions about the physical justification for specific functional forms and the robustness of the derived conclusions. Variations in the choice of shape function can significantly affect assessments of energy conditions, stability analyses, and traversal characteristics \cite{flamm1916comments, visser1995lorentzian}. In recognition of these sensitivities, recent literature has emphasized the necessity for more principled methodologies for shape-function determination—frameworks that integrate mathematical tractability with a deeper physical rationale regarding the interplay between geometry, effective stress–energy, and traversability requisites \cite{zubair2016static,yousaf2017static}.

In this context, Padé approximants represent an effective strategy for the systematic refinement of wormhole shape functions, thereby providing enhanced analytical control \cite{capozziello2020high,gruber2014cosmographic}. The Padé approximants have proven to be effective for modeling wormholes. Preliminary investigations indicate that Padé-based shape functions can reproduce throat profiles with reduced parameter sets, inherently satisfy the flare-out criterion over extended radial intervals, and improve numerical stability in field-equation integrations \cite{saleem2023traversable, capozziello2021traversable,sultan2025pade}. However, the applicability of Padé approximants as a suitable framework for generating sets of shape functions remains an open question, as they can exhibit unpredictable behavior \cite{baker1961pade}. Thus, in this work, we will investigate the limitations of using Padé approximants to model novel TWH shape functions, outlining their strengths and weaknesses through a comprehensive analysis of their applications.

This paper is organized as follows: In Section 2.1 we establish the foundation of a TWH and describe the properties of an adequate shape function. Section 2.2 introduces the Padé approximants. Section 3 is devoted for discuss and modelling of generic shape functions using differents orders of Padé approximants: (3.1) $[1/0]$-order, (3.2) $[0/1]$-order, (3.3) $[1/1]$-order and (3.4) $[L/M]$-order. Finally, in section 4, we state our conclusions.

\section{Theory}

\subsection{The traversable wormholes}

The static, spherically symmetric wormhole metric proposed by Morris and Thorne \cite{morris1988wormholes} can be expressed as follows: 
\begin{equation}\label{EFE}
ds^2 = -e^{2\Phi(r)}\,dt^2 + \frac{dr^2}{1 - \frac{b(r)}{r}} + r^2\bigl(d\theta^2 + \sin^2\theta\,d\phi^2\bigr)\,,
\end{equation}
where \(t\) is the time coordinate, \(r\) the radial coordinate, and \((\theta,\phi)\) the usual angular coordinates. The function \(\Phi(r)\), called the \emph{redshift function}, controls gravitational time dilation and must remain finite everywhere to avoid event horizons. The \emph{shape function} \(b(r)\) determines the spatial geometry of the wormhole: its behavior with \(r\) fixes the throat’s size and the embedding of the spatial slice.

We assume \(r\in [r_0,\infty)\), where \(r_0\) represents the throat radius, and we impose the following geometric requirements on \(b(r)\):
\begin{itemize}
    \item $b(r_0) = r_0$,
    \item $\frac{b(r) - b'(r)r}{b^2} > 0$,
    \item $b'(r_0) - 1 \leq 0$,
    \item $\frac{b(r)}{r} < 1$ for all $r > r_0$,
    \item $\frac{b(r)}{r} \rightarrow 0$ as $r \rightarrow \infty$.
\end{itemize}

An in-depth analysis of these conditions is given as follows. To begin with, $b(r_0)=r_0$ establishes the wormhole throat at $r_0$, ensuring the radial coordinate reaches a smooth minimum without any cusps. The flare-out condition $\bigl(b(r)-b'(r)r\bigr)/b^2>0$ ensures outgoing curves at the throat for spatial slices, necessitating a local breach of the null energy condition. Furthermore, $b'(r_0)\le1$ restricts the throat's expansion rate, maintaining manageable tidal accelerations for travelers. Requiring $b(r)/r<1$ for $r>r_0$ upholds the positivity of the radial metric component, keeping the Lorentzian signature intact and preventing external event horizons. Additionally, $b(r)/r\to0$ as $r\to\infty$ enforces asymptotic flatness, allowing spacetime to approximate Minkowski space at great distances, where gravitational influences normalize. Lastly, ensuring the redshift function $\Phi(r)$ remains finite for $r\ge r_0$ prevents event horizon formation, facilitating bidirectional traversal. Adhering to these requirements results in a nonsingular, traversable wormhole geometry sans event horizons.

\subsection{The Padé approximants}

Consider a function $b(r)$ represented by a power series such that
\begin{equation}\label{series1}
    b(r) = \sum_{n=0}^L b_n \; r^n,
\end{equation}
where $L$ need not be finite, and $b_n$ are the series coefficients. The variable $r$ often plays the role of a perturbative parameter. In perturbation theory, this parameter is introduced so that successive truncations yield a sequence of more tractable problems. We define the Padé approximants as a rational function given by the quotient of two polynomials. The degrees of the numerator and denominator polynomials determine the mixed degree of the approximant. Accordingly, we denote a Padé approximant of degree $(L,M)$ by $[L/M](r)$, where $L$ and $M$ are the degrees of the numerator and denominator polynomials, respectively. That is,
\begin{equation}\label{pade1}
    [L/M](r) = \frac{\Delta_L(r)}{\Gamma_M(r)}, \quad L,M = 0,1,2,3,\dots,
\end{equation}
with $\Delta_L(r)$ being a polynomial of degree $L$,
\begin{equation}
    \Delta_L(r) = \delta_0 + \delta_1 r + \delta_2 r^2 + \dots + \delta_L r^L,
\end{equation}

and $\Gamma_M(r)$ a polynomial of degree $M$,
\begin{equation}
    \Gamma_M(r) = 1 + \gamma_1 r + \gamma_2 r^2 + \dots + \gamma_M r^M,
\end{equation}

where, without loss of generality, we have adopted Baker's normalization condition \cite{baker1961pade} and set the constant term of the denominator to one.

Padé approximants are based on the terms of an underlying power series. Thus, for them to be well-defined and applicable, they must be associated with a specific power series. This power series dictates the values of the coefficients $\delta_i$ and $\gamma_j$. The relationship between the power series \eqref{series1} and the Padé approximant \eqref{pade1} is given by
\begin{equation}\label{matching}
    b(r) - [L/M](r) = \mathcal{O}(r^{L+M+1}).
\end{equation}

Equation \eqref{matching} ensures that $b(r)$ and $[L/M](r)$ are in agreement up to terms of order $r^{L+M}$. In other words, to determine the coefficients $\delta_i$ and $\gamma_j$ uniquely in terms of the series coefficients $b_n$, one requires $L+M+1$ terms of the series \eqref{series1}.

To compute $\delta_i$ and $\gamma_j$, we rewrite \eqref{matching} as
\begin{equation}
    b(r) - \frac{\Delta_L(r)}{\Gamma_M(r)} = \mathcal{O}(r^{L+M+1}),
\end{equation}
from which it follows that
\begin{equation}
    \Delta_L(r) = b(r)\,\Gamma_M(r) + \mathcal{O}(r^{L+M+1}),
\end{equation}
where $\mathcal{O}(r^{L+M+1})$ denotes terms of order $r^{L+M+1}\Gamma_M(r)$. Explicitly writing out the series coefficients, we have
\begin{equation*}
    \begin{split}
        \delta_0 +&\;\delta_1 r + \delta_2 r^2 + \dots + \delta_L r^L = \\
        &(b_0 + b_1 r + b_2 r^2 + \dots)\,\bigl(1 + \gamma_1 r + \gamma_2 r^2 + \dots + \gamma_M r^M\bigr) + \\
        &\hspace{6.8 cm} \mathcal{O}(r^{L+M+1}).
    \end{split}
\end{equation*}
Neglecting terms of order $r^{L+M+1}$ and equating like powers of $r$, one obtains the linear system
\begin{equation}\label{sispade}
    \begin{aligned}
        \delta_0 &= b_0, \\
        \delta_1 &= b_1 + b_0\gamma_1, \\
        \delta_2 &= b_2 + b_1\gamma_1 + b_0\gamma_2, \\
        &\vdots \\
        \delta_L &= b_L + b_{L-1}\gamma_1 + \dots + b_0\gamma_L, \\
        0 &= b_{L+1} + b_L\gamma_1 + \dots + b_{L+1-M}\gamma_M, \\
        0 &= b_{L+2} + b_{L+1}\gamma_1 + \dots + b_{L+2-M}\gamma_M, \\
        &\vdots \\
        0 &= b_{L+M} + b_{L+M-1}\gamma_1 + \dots + b_L\gamma_M,
    \end{aligned}
\end{equation}
with the conventions $b_n\equiv 0$ for $n<0$ and $\gamma_j\equiv 0$ for $j>M$. The structure of system \eqref{sispade} indicates that to determine the coefficients $\delta_i$ and $\gamma_j$, one needs only the first $L+1$ equations and the last $M$ equations. For instance, to compute the coefficients of the approximant $[1/1](r)$, we assemble the first $L+1=2$ equations and the final $M=1$ equation from \eqref{sispade}:
\begin{equation}
\begin{cases}
    \delta_0 = b_0, \\
    \delta_1 = b_1 + b_0\gamma_1, \\
    0 = b_2 + b_1\gamma_1.
\end{cases}
\end{equation}

It immediately follows that
\begin{equation}
    \delta_0 = b_0, \quad \gamma_1 = -\frac{b_2}{b_1}, \quad \delta_1 = b_1 - b_0\frac{b_2}{b_1}.
\end{equation}

To utilize Padé approximants as effective shape functions, they must adhere to the geometric constraints outlined in the previous section. There are two primary applications of Padé approximants for this purpose. First, Padé approximants can be employed to transform an inadequate shape function into a suitable one \cite{capozziello2021traversable}. In this scenario, Padé approximants can generate a new family of viable shape functions from a generic function. Another advantage of Padé approximants is their particularly integrable format, which is useful for replacing complicated shape functions for an equivalent simple rational function \cite{sultan2025pade}.

\section{Applications}

\subsection{The $[1/0]$-order Padé approximant}

There are two main methods to utilize the $[1/0]$-order Padé approximant for the shape function $b(r)$. The first method employs $b(r)/r$ as the original function. However, this approximation is clearly not valid once $[1/0]$ takes the form of a first-order Taylor approximation, $1+\beta'(r-r_0)$, with $\beta' = \frac{b'(r_0)-1}{r_0}$ \cite{capozziello2021traversable}. Thus, for $\frac{b(r)}{r} \rightarrow \infty$ as $r\rightarrow \infty$, which violate asymptotic flatness. The second method uses the approximation as
\begin{equation}\label{firsttaylor}
    b(r) = \alpha + \beta(r - r_0),
\end{equation}
with $\alpha = r_0$ and $\beta = b'(r_0)$.

A viable shape function must satisfy the geometric constraints outlined in the previously. In this section, we will analyze each constraint for the $[1/0]$-order Padé approximant.

The first constraint, $b(r_0) = r_0$, is satisfied for (\ref{firsttaylor}) since $\alpha = r_0$. The second constraint, $\frac{b(r) - b'(r)r}{b^2} > 0$, is the ``flare-out'' condition, ensuring the outward expansion of the TWH throat. Differentiating (\ref{firsttaylor}) yields $b'(r) = b'(r_0)$, indicating that $b'(r)$ is constant. This implies
\begin{equation*}
    b(r) - b'(r)r = r_0\left[1 - b'(r_0)\right].
\end{equation*}
The expression above relates the ``flare-out'' condition, on the left side, with the third condition, $b(r_0)-1\le 0$, on the right side. The third condition is not a function of $r$, which implies that it can be more easily satisfied for an arbitrary function approximated by the $[1/0]$-order Padé approximant. Thus, when $b(r_0)-1\le 0$ is satisfied, the ``flare-out'' condition is also satisfied.

\begin{figure}[h]
    \centering 
    \includegraphics[width=0.48\textwidth]{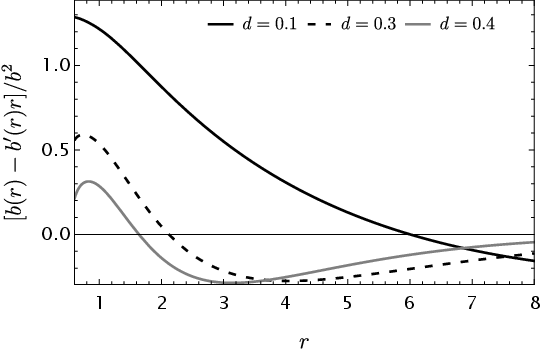}	
    \caption{The ```flare-out" condition for (\ref{original}) as a function of $r$ with $r_0 = 0.6$ and $d = 0.1, 0.3$ and $0.4$.}
    \label{figbor}
\end{figure}

\begin{figure}[h]
    \centering 
    \includegraphics[width=0.48\textwidth]{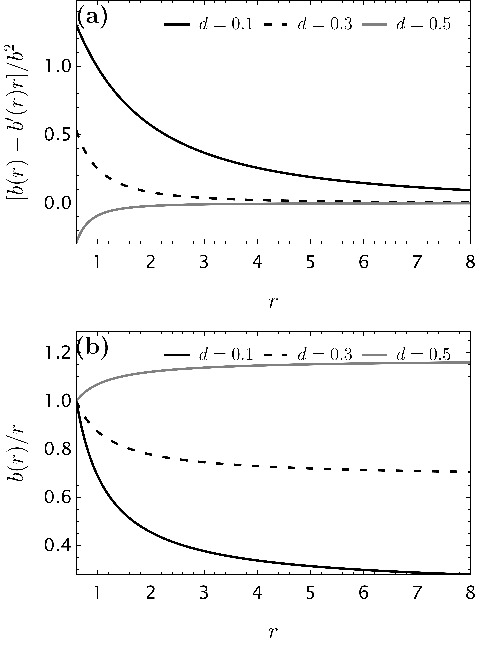}	
    \caption{(a) The ``flare-out" condition and (b) the fourth condition for  $[1/0]$-order Padé approximant of (\ref{original}), with $b(r)$ as the approximated function, as a function of $r$ for $d \le 0.43$ ($d=0.1$ and $0.3$) and $d>0.43$ ($d=0.5$) and $r_0 = 0.6$.}
    \label{figbr}
\end{figure}

To illustrate the preceding discussion, we can consider a candidate for the shape function:
\begin{equation}\label{original}
    b(r) = r_0\left[\arctan\left(\frac{r}{r_0}-1\right) + \sinh{\left(\frac{r}{r_0}\right)}\right]^d,
\end{equation}
where $d$ is the real-valued constant and $r_0$ is the throat of the TWH. The Eq.(\ref{original}) represents an inadequate shape function. At first glance, we observe that $b(r_0)$ is not equal $r_0$. In addition, the Fig.~\ref{figbor} shows negative values for the ``flare-out`" condition, indicating that this constraint is not satisfied. On the other hand, the third condition limits the range of validity with $d\le0.431$. Therefore, (\ref{original}) is not an appropriate shape function.   
To produce an adequate shape function from (\ref{original}) we utilize the $[1/0]$-order Padé approximant,which yields  
\begin{equation}\label{firstorder}
    b(r) = r_0 +2.15d(r_0 - r)1.18^d.
\end{equation}
The $[1/0]$-order Padé approximant, or first-order Taylor expansion, creates a new function that satisfies all previously discussed constraints, Fig~\ref{figbr}. However, the third condition imposes a restriction on the values of $d$, and this restriction persists for any $[L/M]$-order Padé approximant. This occurs because the restriction is produced by $b'(r_0)$, and the derivative at the point $r_0$ is the same for all Padé approximants generated for (\ref{original}). In addition, we can verify that violation of the third constraint also violates the fourth condition, $b(r)/r < 1$ for $r > r_0$, but adherence ensures compliance with the first four constraints. The valid, and non-valid, range are illustrated in Fig.~\ref{figbr}, where the the ``flare-out" and $b(r)/r<1$ are showed as a function of $r$ for values of $d\le0.43$ and $d>0.43$. In addition, it is necessary $d >0$, for a $b(r) >0$. Thus, the $[1/0]$-order Padé approximant, until now, produced an adequate shape function, but it is not capable of to remove any restriction in your parameters produced by $b'(r_0)$. 

The discussion in the preceding paragraphs is crucial for assessing the suitability of the $[1/0]$-order Padé approximant as a shape function for a TWH. We have demonstrated that if the $\alpha$ term is replaced by $r_0$ and $b'(r_0) < 1$ are satisfied, the first four geometric restrictions are fulfilled. Now, more attention must be paid to the final constraint $\frac{b(r)}{r} \rightarrow 0$ as $r \rightarrow \infty$, which requires careful analysis. Multiplying (\ref{firsttaylor}) by $1/r$ and taking $r \rightarrow \infty$ yields
\begin{equation}
    \frac{b(r)}{r} \rightarrow \beta.
\end{equation}
Thus, the constraint $\frac{b(r)}{r} \rightarrow 0$ requires, at first glance, $\beta = 0$. However, this condition can be relaxed, as the wormhole generates a topological defect similar to a global monopole on large scales.

\begin{figure}[h]
    \centering 
    \includegraphics[width=0.48\textwidth]{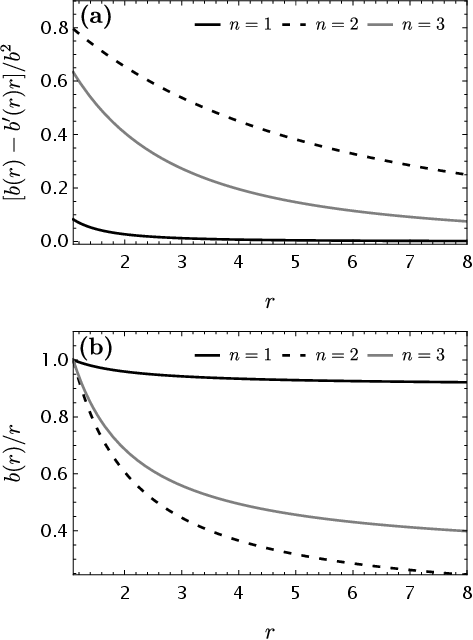}	
    \caption{(a) The ``flare-out" condition and (b) the fourth condition for  $[1/0]$-order Padé approximant of a generic shape function $b(r)$, in the format of \eqref{firsttaylor}, as a function of $r$, with $r_0 = 1.1$ and $b'(r_0) = 1/nr_0$, with $n=1,2,3$.}
    \label{fig2}
\end{figure}

To generalize our approach, we will consider $\beta = 1/nr_0$ with $n\ge 1$ to satisfy $b'(r_0) - 1 \le 0$, and one will plot in Fig.\ref{fig2} the main geometric constraints for a generic $b(r)$ as the form of (\ref{firsttaylor}). The other constraints are not shown as they are straightforward to verify and do not depend on $r$, except for the last constraint, which has already been examined.  The figure demonstrates that both the ``flare-out" condition and $b(r)/r < 0$ are satisfied for any $r$. This indicates that any arbitrary function approximated by $[1/0]$-order Padé approximant, in terms of (\ref{firsttaylor}), can be a shape function of a TWH. These results agree with \cite{Cataldo2017}, which analyzed the viability of a generic linear shape function. 

\subsection{The $[0/1]$-order Padé approximant}

Now, we will investigate the use of the $[0/1]$-order Padé approximant as a candidate for the shape function of a TWH. This approximation takes the form of a rational function, expressed as
\begin{equation}\label{pade01}
    \chi(r) = \frac{\gamma}{1+\delta r},
\end{equation}
where $\gamma$ and $\delta$ are parameters determined by the Taylor series expansion of a generic shape function candidate $b(r)$, as mentioned previously. The $[0/1]$-order Padé approximant parameters are defined by matching two terms of the Taylor series of $b(r)$. The function $\chi(r)$ represents the approximated function by Padé approximant, and, as demonstrated for the $[1/0]$-order Padé approximant, there are two main formats for it. The first format utilizes the shape function itself, $b(r)$, while the second uses the rational function $b(r)/r$ for approximation. Firstly, we will examine $\chi(r)$ approximated for $b(r)$.

When we approximate $b(r)$ for a $[0/1]$-order Padé approximant, we have
\begin{equation}\label{pade01_3}
    b(r) = r_0\left(\frac{1+\delta' r_0}{1+\delta' r}\right),
\end{equation}
with $\delta' = - \frac{b'(r_0)}{r_0(1+b'(r_0))}$. The $\delta'$ was calculated to satisfy $b(r_0) = r_0$. Observing (\ref{pade01_3}), it is easy to verify that $\frac{b(r)}{r} \rightarrow 0$ as $r\rightarrow \infty$ and $b(r)/r$ for all $r > r_0$ are both satisfied. Thus, our effort will focus on ``flare-out'' condition and $b'(r_0)-1<0$.

To analyze the ``flare-out'' condition, we compute the derivative  of (\ref{pade01_3}) and assume $b(r)$ real-valued. Note that for $b'(r_0) > 0$ we have $\delta' <0$. In this scenario, the application of the ``flare-out'' condition determines 
\begin{equation}\label{limitr}
    r<r_0\xi,
\end{equation}
with $\xi = \frac{1+b'(r_0)}{2b(r_0)}$ and $\delta' < - 1/r_0$. This inequality defines the validity range of $b(r)$ as a shape function. However, (\ref{limitr}) limits the range of $r$, which is problematic since $r$ is defined at infinity. Furthermore, for $b'(r_0) > 0$, $\delta' < - 1/r_0$ is not possible. This condition is only valid for $b'(r_0) < -1$, which implies $r>r_0\xi$, with $0<\xi<1$. Thus, $r$ can go to infinity without restriction. Then, if $b'(r_0)<-1$, all geometric constraints are satisfied for \eqref{pade01_3}. 

\begin{figure}[h]
    \centering 
    \includegraphics[width=0.48\textwidth]{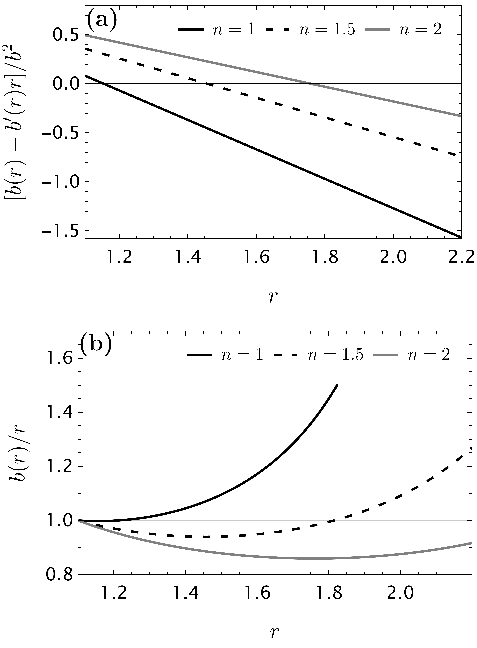}	
    \caption{(a) The ``flare-out" condition and (b) the fourth condition for  $[0/1]$-order Padé approximant of a generic shape function $b(r)$, in the format of \eqref{pade01_3}, as a function of $r$, with $r_0 = 1.1$ and $b'(r_0) = 1/nr_0 > 0$, with $n=1,1.5,2$.}
    \label{fig3}
\end{figure}

To illustrate the effects of $b'(r_0)>0$ we refer to Fig.~\ref{fig3}, where we plot the $[0/1]$-order Padé approximant for a generic $b(r)$, represented by Eq.(\ref{pade01_3}). In Fig.~\ref{fig3}(a), the "flare-out" condition is not satisfied for $r\ge r_0\xi$. In Fig.\ref{fig3}(b), $b(r)/r$ exceeds $1$, diverging for $r=-1/\delta'$. The divergent behavior arises because Eq.(\ref{pade01_3}) has a pole at $r=-1/\delta'$. The Padé approximants can exhibit artificial poles \cite{stahl1998spurious,beardon1968location,suetin2002pade}. To analyze the position of this pole, consider $b'(r_0)=n$ with $m=n/(n+1)$, yielding $\delta'=-m/r_0$, which leads to a pole at $r=r_0/m$. If $n>0$ then $0<m<1$, resulting in a pole for $r>r_0$, which clarifies the behavior shown in Fig.\ref{fig3}(b).

\begin{figure}[h]
    \centering 
    \includegraphics[width=0.48\textwidth]{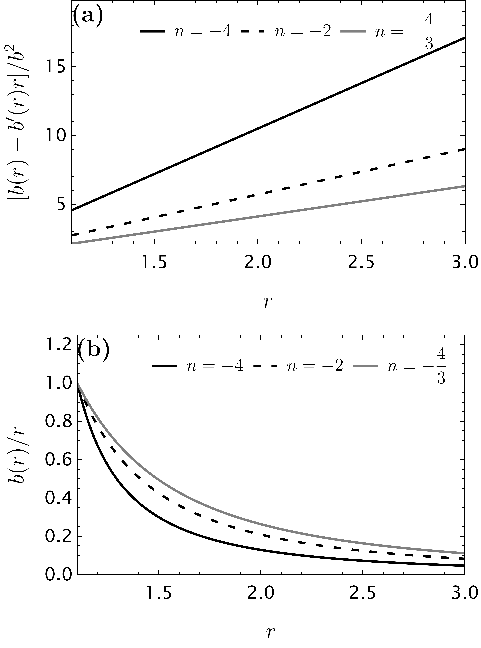}	
    \caption{(a) The ``flare-out" condition and (b) the fourth condition for  $[0/1]$-order Padé approximant of a generic shape function $b(r)$, in the format of \eqref{pade01_3}, as a function of $r$, with $r_0 = 1.1$ and $b'(r_0) = n<0$, with $n=-4,-2,-\frac{4}{3}$.}
    \label{fig4}
\end{figure}

Although the Eq.(\ref{pade01_3}) does not satisfy the geometric constraints for $b'(r_0)>0$, all restrictions are met for $b'(r_0) <-1$. This is illustrated in Fig.~\ref{fig4}. This is achieved because when $b'(r_0)=n<-1$ we have  $m > 1$, which moves the pole into the region $r<r_0$. The new pole position of (\ref{pade01_3}) transforms the $[0/1]$-order Padé approximation of $b(r)$ into an appropriate shape function, fulfilling all geometric constraints.The linear behavior of Fig.~\ref{fig4}(a) is explained by (\ref{limitr}).

\begin{figure}[h]
    \centering 
    \includegraphics[width=0.48\textwidth]{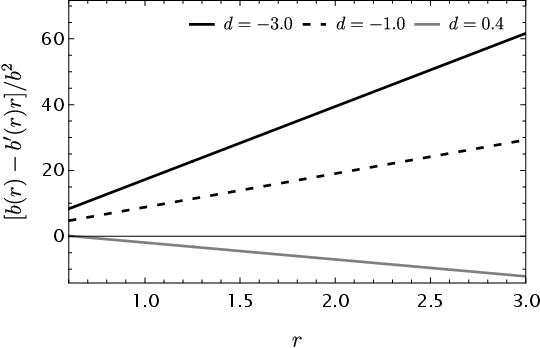}	
    \caption{The ``flare-out" condition for  $[0/1]$-order Padé approximant of (\ref{original}), with $b(r)$ as the approximated function, in the format of \eqref{pade01_3}, as a function of $r$, with $r_0 = 0.6$ and $d =-3.0,-1.0$ and $0.4$.}
    \label{fig8}
\end{figure}

To illustrate the application of this new family of shape function candidates, we can take (\ref{original}) approximated by (\ref{pade01_3}), which is given by
\begin{equation}\label{pade01orib}
   b(r)= \frac{0.462117 r_0^2}{0.462117r_0-1.1752^d d( r-r_0)},
\end{equation}
with $-24.6 < d < -0.50$. We plot (\ref{pade01orib}) in  Fig.~\ref{fig8}, where we present the ``flare-out" condition for the $[0/1]$-order Padé approximant. In Fig.~\ref{fig8}, we observe the effects of the signal of $b'(r_0)$ on the behavior of the approximation. By fixing $b'(r_0) < -1$, we establish a range for $d$, specifically $-24.6 < d < -0.50$. This range results in negative values for $b'(r)$.  When we have $d=0.4 (> -0.50)$, we obtain $b'(r_0)>0$, which violates the valid range of (\ref{pade01orib}), as we can observe in Fig.~\ref{fig8} with the negative values of the``flare-out" condition. It is important to note that $b'(r_0)$ of (\ref{original}) is independent of $r_0$. However, for other candidate shape functions, there may be a dependence on $r_0$ \cite{capozziello2021traversable,sultan2025pade}.

For now, we will examine whether the $[0/1]$-order Padé approximant of $b(r)/r$ is suitable for representing a shape function of a TWH. Following the previous procedures, we can express the $\chi(r)$ as  
\begin{equation}\label{pade01_4}
    \frac{b(r)}{r} = \frac{1+\delta r_0}{1+\delta r},
\end{equation}
with $\delta = \frac{1-b'(r_0)}{r_0b'(r_0)}$. The $\delta$ was calculated to satisfy $b(r_0) = r_0$. The case of (\ref{pade01_4}) is unrestricted regarding parameter values, except for $b'(r_0)-1<0$. The third condition is sufficient for (\ref{pade01_4}) to satisfy all geometric constraints. The Eq.(\ref{pade01_4}) also indicates a pole in $r=1/\delta$. Thus, the previous analysis can be conducted here, considering $\delta = k/r_0$, $k=\frac{1-b'(r_0)}{b'(r_0)}$, and the position of the pole given for $r=-r_0/k$. By analyzing $k$ and the restriction $b'(r_0) -1<0$, we can consider $b'(r_0) >0$ or $b'(r_0)<0$. For $b'(r_0)>0$, we have $k>0$ with the pole at $r<0$, which is located in a non-physical region. For $b'(r_0) <0$, we have $k<-1$, which moves the pole for $r<r_0$, placing it in a region that is not relevant for a TWH. Thus, the pole of (\ref{pade01_4}) never resides in a physically relevant region.

\begin{figure}[h]
    \centering 
    \includegraphics[width=0.48\textwidth]{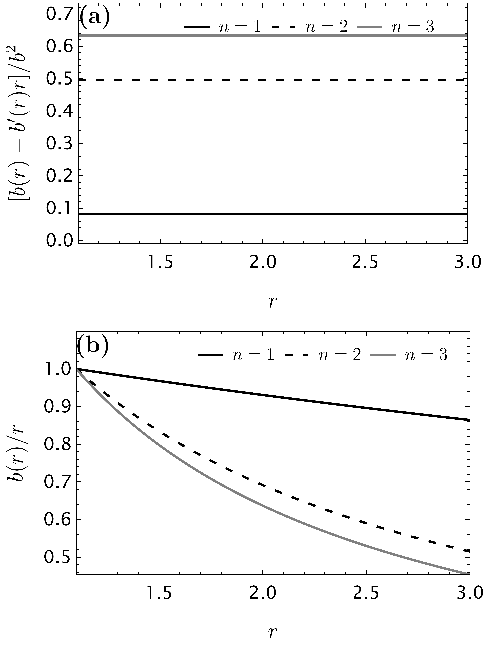}	
    \caption{(a) The ``flare-out" condition and (b) the fourth condition for  $[0/1]$-order Padé approximant of a generic $b(r)/r$, in the format of \eqref{pade01_4}, as a function of $r$, with $r_0 = 1.1$ and $b'(r_0) = 1/nr_0$, with $n=1,2,3$.}
    \label{fig5}
\end{figure}

In Fig.~\ref{fig5} we illustrate the adherence of (\ref{pade01_4}) to the ``flare-out" and $b(r)/r<1$ conditions when $b'(r_0)=1/nr_0$, with $n \ge 1$.  The Fig.~\ref{fig5}(a) shows that the ``flare-out" condition keeps constant and equal $\frac{1-b'(r_0)}{r_0}$. Thus, if $b'(r_0)-1<0$, the ``flare-out" condition is automatically satisfied. 

To exemplify the format of (\ref{original}) when approximated by (\ref{pade01_4}), we take a new shape function given by
\begin{equation}\label{pade01orib2}
   b(r)= \frac{0.462117 r_0}{d e^{0.161439d}r_0-(1.1752^d d - 0.462117) r}r,
\end{equation}
with $d < 0.43$. In Fig.~\ref{fig7}, we present the ``flare-out" condition for the $[0/1]$-order Padé approximant for $b(r)/r$, with $b(r)$ replaced by (\ref{original}). It is important to note that the range for $d$, as presented for the $[1/0]$-order Padé approximant, persists for the $[0/1]$ approximant, which can be observed when we consider $d>0.43$. However, the restrictions imposed on the use of $b(r)$ as the approximated function instead of $b(r)/r$, $-24.6 < d < -0.50$, are no longer valid. This explains the positive values of ``flare-out" for $d=0.4$, in Fig.~\ref{fig7} . Furthermore, it is important to note that although there exists a limited range for $d$, this does not impose any constraints on the range of $r$. 
\begin{figure}[h]
    \centering 
    \includegraphics[width=0.5\textwidth]{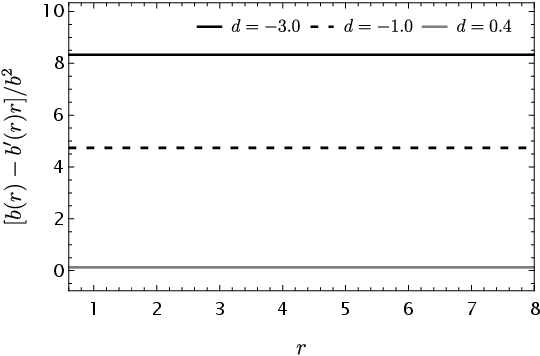}	
    \caption{The ``flare-out" condition for  $[0/1]$-order Padé approximant of (\ref{original}), with $b(r)/r$ as the approximated function, in the format of \eqref{pade01_4}, as a function of $r$,  with $r_0 = 0.6$ and $d =-3.0,-1.0$ and $0.4$.}
    \label{fig7}
\end{figure}

\subsection{The $[1/1]$-order Padé approximant}

The $[1/1]$-order, or first-order, Padé approximant is a rational function that consists of linear functions in both the numerator and  denominator. In this context, the first-order Padé approximant can be expressed in two primary forms, as discussed in the previous sections concerning the lowest-order Padé approximants. Specifically, both $b(r)$ and $b(r)/r$ can be employed as functions to generate a first-order Padé approximant as follows: 
\begin{equation}\label{pade11}
    \upsilon(r) = \frac{\alpha+\beta r}{1+\gamma r},
\end{equation}
where $\upsilon(r)$ can be $b(r)$ or $b(r)/r$. For $\upsilon(r) = b(r)/r$, we have
\begin{subequations}\label{parameters}
\begin{equation}\label{parameter_a}
    \alpha = 1-(1+r_0\gamma)(b'(r_0)-1),
\end{equation}
\begin{equation}\label{parameter_b}
    \beta = \frac{b'(r_0)(1+r_0\gamma)-1}{r_0},
\end{equation}
and
\begin{equation}\label{parameter_g}
    \gamma = \frac{2(b'(r_0)-1)-r_0b''(r_0)}{r_0^2 b''(r_0)},
\end{equation}
\end{subequations}
where $b''(r_0)$ is the second derivative of $b(r)$ in $r_0$. The parameters in Eqs.\ref{parameters}[a-c] were computed to satisfy $b(r_0)=r_0$.

\begin{figure}[h]
    \centering 
    \includegraphics[width=0.48\textwidth]{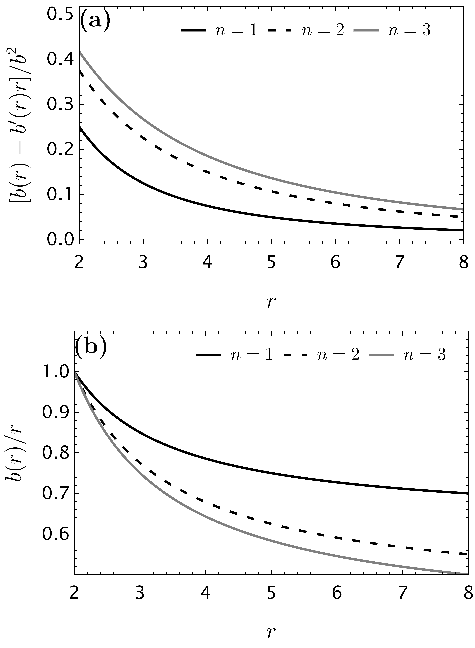}	
    \caption{(a) The ``flare-out" condition and (b) the fourth condition for $[1/1]$-order Padé approximant of a generic $b(r)/r$, in the format of \eqref{pade01_4}, as a function of $r$, with $n=1,2,3$, and $\gamma = -r_0$.}
    \label{fig6}
\end{figure}

The first-order Padé approximant can be a shape function of a TWH if the geometric constraints are  satisfied.  The first constraint was fulfilled when we defined $\alpha, \beta$ and $\gamma$ in \eqref{parameters}[a-c]. For the ``flare-out" condition, it is required that $\beta < \gamma \alpha$. This requirement implies $b'(r_0)-1<0$. Thus, as the third condition is satisfied, we have $\alpha > 1$ for $1+r_0\gamma > 0$ and $\alpha < 1$ for $1+r_0\gamma < 0$. The derivative of $b(r_0)$ in $r_0$, $b'(r_0)$, can take on both positive and negative values. For positive values, we have $1+r_0\gamma >0$, which implies $\beta>-1/r_0$, or $1+r_0\gamma <0$, which implies $\beta<-1/r_0$. By analyzing $b(r)/r < 1$ as $r>r_0$, we observe that if the previous constraints are met, this condition also holds true. However, when we consider $r\rightarrow0$, we find $b(r)/r \rightarrow \beta/\gamma$, a nonzero value, analogous to the [1/0]-order Padé approximant, which indicates a topological defect similar to a global monopole on large scales. It is important to note that $b(r)/r$ can take on negative values when we have $b'(r_0)<1/(1+r_0\gamma)$. However, $1+r_0\gamma$ can assume both negative and positive values. To determine which signal is physically valid, we need to analyze the poles of (\ref{pade11}). The pole of the first-order Padé is located at $r_p=-1/\gamma$. Thus, when we take $1+r_0\gamma<0$, the pole appears for $r_p<r_0$. Consequently, the physically valid choice for $\gamma$ is $\gamma < -1/r_0$. This analysis demonstrates that for any negative values of $b'(r_0)$, $b(r)/r$ is always negative. In Fig.~\ref{fig6}, we illustrate the ``flare-out" condition in Fig.~\ref{fig6}(a) and $b(r)/r<1$ in Fig.~\ref{fig6}(b), both for $b'(r_0)<1$ and $\gamma <-1/r_0$.

Although the $[1/0]$-order Padé approximant can serve as a suitable shape function for a TWH, the mentioned restrictions must be observed. The numerous constraints arising from the presence of $b''(r_0)$ in your general formulation (\ref{pade11}) limit its applicability. To illustrate this situation, we can refer to the first-order Padé approximation of (\ref{original}). Naturally, the geometric constraints of a TWH impose a specific range for $d$, i.e. $d<0.43$. However, since $\gamma < -1/r_0$ is a necessary condition for the suitability of  (\ref{pade11}) as a shape function, it must be satisfied. Thus, when $\gamma < -1/r_0$, we have $\alpha < 1$. This last restriction imposes an additional range for $d$, $d>0.43$. This new range is incompatible with the previous one. Consequently, the first-order Padé approximant of (\ref{original}), using $b(r)/r$ as the approximate function, is not an adequate shape function. This behavior can be explained partly by the additional restrictions imposed for introducing the second derivative in \eqref{pade01_4}, and partly by the Padé approximant's ability to converge rapidly to the original function (\ref{original}), which is not a suitable shape function for a TWH. 

Although the first-order Padé using $b(r)/r$ is not suitable for generating a valid shape function for \eqref{original}, we can attempt to directly replace $\upsilon(r)$ with $b(r)$. This approach is simpler and does not impose the additional restriction $\alpha <1$, which has now been replaced by $\alpha > 0$. The new restriction does not impose an additional range for $d$, maintaining only $d<0.43$. Thus, for the \eqref{original}, the direct approximation of $b(r)$ is more effective. The new shape function obtained is given by
\begin{equation}\label{originalpade11}
b(r)= \frac{r_0\left[2.350\,r_0 + 1.1752^d\,(2+2.5431\,d)(r-r_0)\right]}
{2.3504\,r_0+(2-2.5431\,d)(r-r_0)},
\end{equation}
with $d<0.43$. The straightforward format of the \eqref{originalpade11} demonstrates that the Padé approximants provide a strong framework for transforming a complex shape function into a simpler rational shape function, which is generally more conducive to integration. Padé approximants offer a robust framework for transforming even a complex or initially unsuitable shape function into a simplified, well-suited rational function. Despite this simplification, the resulting rational function preserves sufficient complexity to maintain a rich structure.

\begin{figure}[h]
    \centering 
    \includegraphics[width=0.49\textwidth]{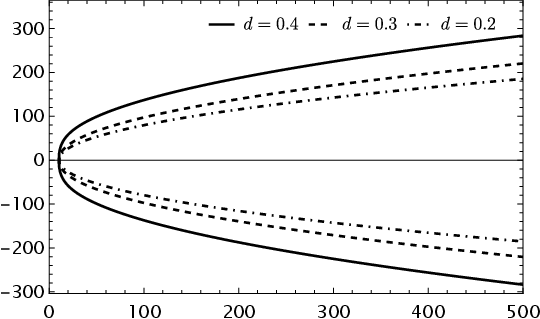}	
    \caption{Profiles of the embedding diagrams for the wormhole solution in the the [1,1]-order Padé approximant, considering some allowed values of $d$. The throat radius is set $r_0=10$.}
    \label{profiles}
\end{figure}

In Fig. \ref{profiles}, we show the wormhole profiles from the embedding diagrams based on the solution given by Eq. (\ref{originalpade11}), for varying values of \(d\) within the allowed range. It is important to note that the spatial curvature near the throat is more pronounced for lower values of \(d\), with these wormholes approaching asymptotic flatness more rapidly.

\subsection{The $[L/M]$-order Padé approximat}

We define the high-order Padé approximants as rational functions with $L>1$ and $M>1$. The use of high-order approximants for modeling  shape functions presents several significant challenges. A primary limitation arises from the increased probability of artificial poles emerging within the physical domain $r \geq r_0$ as the order of the approximant increases. These poles directly conflict with the fundamental geometric requirement $b(r)/r < 1$, and disrupt the necessary asymptotic behavior $b(r)/r \to 0$ as $r \to \infty$. Even when poles are technically situated outside the physical range, their proximity can introduce numerical instabilities during the integration of the field equations.

The high-order approximants require matching additional Taylor series coefficients, which introduce more parameters $\delta_i$ and $\gamma_j$ associated with higher derivatives $b^{n}(r_0)$. While this enhancement improves local approximation near the throat at $r_0$, it results in an overdetermined system where the parameters must satisfy multiple constraints simultaneously. High-order Padé approximants may exhibit unexpected behavior, violating the requirements of wormhole geometry, particularly when the original shape function is not inherently suitable. These challenges make high-order Padé approximants an unpredictable framework for generating appropriate shape functions. To illustrate the limitations of the high-order approximants, we calculated diagonal Padé approximants of various orders for \eqref{original}, using $b(r)$ and $b(r)/r$ as generators of the Taylor series. When we use $b(r)$ directly, the $[2/2]$-order Padé works well; however, the $[3/3]$, $[4/4]$, and $[5/5]$-order Padé approximants exhibit artificial poles for $r> r_0$. When we utilize $b(r)/r$, the $[2/2]$, $[3/3]$ and $[4/4]$-order Padé approximants perform well for $d<-0.5$, but $[3/3]$ presents poles for $d>0$. The $[5/5]$ does not exhibit poles, but it shows a negative "flare-out". This behavior occurs because, starting from the $[4/4]$, the Padé approximants replicate \eqref{original} over an ever-increasing range of $r$. To illustrate this, we plot in Fig.\ref{fig7} the Padé approximants up to $[13/13]$ to visualize the adherence of the approximations to the original function \eqref{original}. We only plot the Padé approximants that do not present poles for $r>r_0$. Thus, the high-order approximants from $[4/4]$ are unsuitable as shape functions. This behavior is specific to \eqref{original}, and other candidates may exhibit different behavior; that is, we have no control over which order the Padé approximants will converge to the original function. Therefore, we cannot assert that the high-order intermediate Padés (approximants that do not exhibit relevant convergence ranges) will be suitable shape functions. Thus, the challenges arising from the presence of poles and high-order derivatives of $b(r)$ can persist across general examples.
\begin{figure}[h]
    \centering 
    \includegraphics[width=0.48\textwidth]{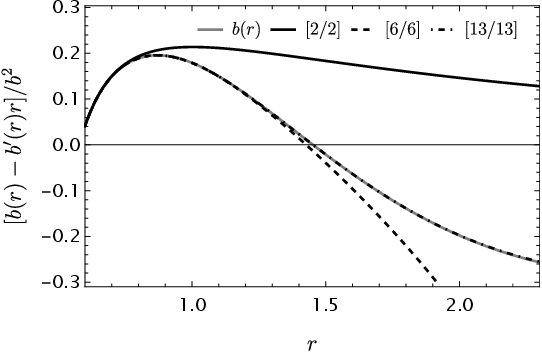}	
    \caption{The ``flare-out" condition, as a function of $r$, for \eqref{original} and differents high-order Padé approximants of \eqref{original}, generates from $b(r)/r$, with $r_0=0.6$ and $d=0.45$.}
    \label{fig7}
\end{figure}

\section{Summary and conclusions}

We demonstrate the usefulness and importance of Padé approximants in the systematic construction and analysis of traversable wormhole shape functions. By employing low-order approximants $[1/0]$, $[0/1]$, and $[1/1]$ to linearize and rationalize candidate profiles, we identify the precise parameter regimes in which the flare-out, throat regularity, and asymptotic flatness conditions are simultaneously satisfied. Furthermore, we show how each approximant transforms complex or non-viable functions into integrable rational forms. While these low-order Padé approximants offer robust analytical control for arbitrary shape functions, our investigation of higher $[L/M]$ orders reveals the emergence of spurious poles and overdetermined constraints that can compromise geometric viability. Moreover, we lack control over when the Padé approximant may exhibit inadequate behavior, as this depends on the nature of the approximated function. In this paper, we focused on the geometric aspects of traversable wormholes; however, further studies on energy conditions are necessary for a comprehensive physical understanding of the shape functions derived from Padé approximants.

\section*{Acknowledgements}
CRM thanks to Conselho Nacional de Desenvolvimento Científico e Tecnológico for the partial support, Grant no. 308268/2021-6.




\bibliographystyle{elsarticle-num} 
\bibliography{example}






\end{document}